# Strong impact of the eddy–current shielding on ferromagnetic resonance response of sub–skin–depth–thick conducting magnetic multilayers

Ivan S. Maksymov[1,*], Zhaoyang Zhang[1,2], Crosby Chang[1], and Mikhail Kostylev[1]

[1]School of Physics, University of Western Australia, 35 Stirling Highway, Crawley WA 6009, Australia

[2]University of Science and Technology of China

*corresponding author, ivan.maksymov@uwa.edu.au

**Abstract:** Exchange–coupled non–magnetic metal (NM) and ferromagnetic metal (FM) multilayers are crucial for microwave magnonic and spintronic devices. These layered materials usually have total thicknesses smaller than the microwave skin depth. By using a stripline broadband ferromagnetic resonance spectroscopy technique, we experimentally demonstrate that the amplitude of the magnetisation precession in the FM layer is strongly diminished by the shielding effect of microwave (6–12 GHz) eddy currents circulating in the NM capping layers.

# 1. Introduction

Magnonics and spintronics are emerging nanotechnologies offering functionalities beyond the current semiconductor technology.[1] Consequently, there is a huge interest in the excitation, detection and control of spin waves at the nano–scale,[2] as well as in a number of spintronic effects such as the spin transfer torque, direct and inverse Spin–Hall effects, and spin pumping.[3,4] A deep understanding of physics of these effects is crucial for the development of novel devices that include but not limited to: magnetic random access memory (MRAM), spin–torque MRAM, spin–torque nano–oscillators,[1,3] frequency–agile left–handed meta-materials[5–10], as well as gas sensors.[11]

The aforementioned devices operate at microwave frequencies and they are made of multilayers consisting of non–magnetic metal (NM) and ferromagnetic metal (FM) thin films. A large amount of literature is devoted to the investigation of magnetisation dynamics and spin transport parameters of such multilayers (e.g., [1–4, 11–16] to cite just a few recent articles).

Permalloy (Py = $Ni_{80}Fe_{20}$) is the material of choice for the FM layer because it exhibits an optimal combination of magnetic properties (e.g., the vanishing magnetic anisotropy and one of the smallest magnetic (Gilbert) losses $\alpha_G$ among ferromagnetic metals).[2,17–19] NM layers are usually made of Ta, Cu, Au, Pt, or Pd.[3,4,11–13] Ta and Cu thin films often act as a seed layer[20,21] or as a capping layer protecting the Py film from oxidation.[22] Pt and Pd layers are used in spintronic devices exploiting the spin pumping and Spin–Hall effects[3,4] as well as in gas sensors.[11] The spin pumping effect in Cu layers is negligible.[23] Hence, multilayers with NM = Cu are often used as reference samples.[15,24]

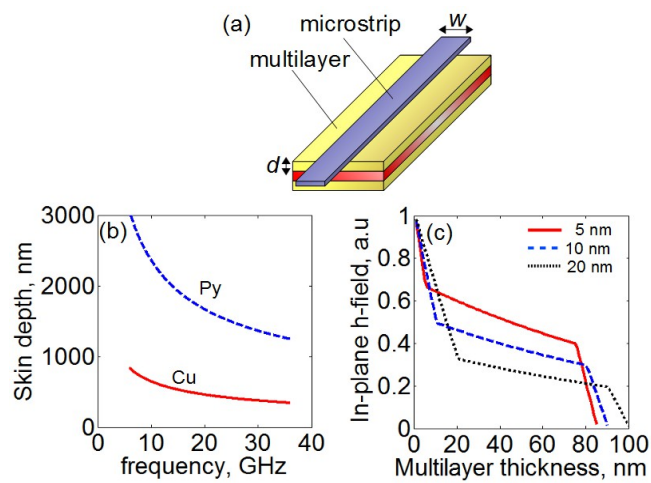

**FIG. 1** (a) Schematic of a NM/FM/NM multilayer facing a microstrip line with the NM capping layer of thickness $d$. (b) Theoretical skin–depth $\delta$ of Cu and Py as a function of the microwave frequency $f$. (c) Numerically simulated profiles of the in-plane component of the microwave magnetic field across the multilayers with different Cu capping layer thicknesses. The total thickness of the multilayer is counted from the side facing the ML. The microwave frequency $f = 12$ GHz and the thickness of the Cu seed layer is 10 nm in all cases.

All the aforementioned applications rely on exposure of the metallic multi–layered materials to a microwave electromagnetic field. Most often this field is incident on just one of the two sample surfaces. The most popular way to expose a planar sample to a microwave radiation is by using broadband stripline ferromagnetic resonance (FMR) method (see, e.g., Refs. [25–29]). In particular, this method is largely used to study the magnetisation dynamics and spin current injection through interfaces (spin pumping and inverse spin Hall effects).

The main part of a broadband–FMR (BFMR) setup is a section of a microstrip line (ML). The multilayer under test sits on top of the ML [Fig. 1(a)]. A microwave current flowing through the ML at a fixed frequency $f$ imposes a microwave (Oersted) magnetic field on the multilayer. The resonance frequency in the multilayer is determined by a slowly scanned frequency $f$ or external static magnetic field $H$. In the latter case, as the value of $H$ is adjusted, the frequency of the natural magnetisation precession resonance eventually equals the frequency of the microwave magnetic field, and significant microwave power absorption occurs.

Although the impact of NM layers on the linewidth ($\Delta H$) of the FMR response is well–known,[14,15,23,30] the effect of their electrical conductivity $\sigma$ has usually been neglected while designing BFMR experiments. The main reason for this is the belief that metal layers thinner than the microwave skin–depth $\delta$ [Fig. 1(b)] do not affect magnetisation dynamics.[31] However, many theoretical analyses have called this assumption into question.[22,32–38] In our Refs. [35, 36], we show theoretically that the in–plane component of the microwave magnetic field vanishes at the far end of conducting multilayers with sub-skin-depth thicknesses [Fig. 1(c)].[39] Most significantly, the amplitude of the microwave magnetic field in the FM layer drops very quickly when the thickness of the capping layer $d$ is increased [compare the central sections of the curves in Fig. 1(c)].

In this work, we present experimental evidence of a strong and adverse effect of the conducting NM (Cu) capping layers on the strength and profile of the FMR response of the underlying Py layer. The eddy currents circulating in the capping layer shield the Py layer from microwave magnetic fields induced by the ML transducer of the BFMR setup. We show that the shielding is very strong even when the thickness of the NM capping layer is well below the microwave skin–depth δ for the NM material.

Although in the following we focus on the stripline measurements, our findings are applicable to a broader class of situations, in which the microwave magnetic field is also incident on one film surface only. This possibility of generalisation was demonstrated in our recent experiment[40] and confirmed theoretically.[41] Indeed, in the conditions of the stripline FMR samples are exposed to a *near* microwave magnetic field of a stripline. However, in Refs. [40, 41] we showed that exposure of metallic ferromagnetic films samples to a *far* field results in the same behaviour. For instance, the response in reflection of the samples exposed to a plane *travelling* electromagnetic wave incident on a sample surface from free space is very similar to the stripline BFMR response. The similarity of the far– and near–field responses is very important, e.g., for experimental characterisation of magnetic meta–materials.[5–10] Note that the far–field and near–field responses are similar, provided the stripline width is large enough as in our experiment. The requirement of a large stripline width ensures the absence of the adverse effect of travelling spin wave contribution to the FMR response[27] and should be fulfilled in any BFMR experiment.

## 2. Experimental results and discussion

We use a macroscopic 0.33 mm–wide ML transducer and place the sample grown on a Si substrate on top of it. As we demonstrated previously,[22] for macroscopic transducers the eddy–current effects should be very strong. We carry out BFMR measurements on magnetron sputter–deposited Si/Cu[10nm]/Py[70nm]/Cu[$d$] multilayers with $d$ = 10, 20, 35 and 70 nm. The capping layer of the samples faces ML [Fig. 1(a)]. We keep the microwave frequency $f$ constant and sweep the magnetic field $H$ applied in the sample plane and along the transducer to record the raw FMR absorption traces. This procedure is repeated for several values of $f$ is in the range 6–12 GHz with the 2 GHz step. To record the traces we use a freshly calibrated microwave network analyser. The measurements are taken at room temperature.

Although in many experimental situations the thickness of the NM capping layers is <10 nm,[3,4,11–13] 10 nm thick capping layers are also often used, e.g., in Pt–YIG (yttrium–iron–garnet) magnetic structures.[42,43] In this work, we would like to address this very important case. Furthermore, the impact of the eddy currents strongly depends on the frequency (see Fig. 5 in Ref. [22]). Therefore, our findings of the effect of the 10 nm–range capping layers in the 6–12 GHz frequency range may be important for experiments at ~40 GHz employing samples with $d = 5$ nm or so.

In order to demonstrate the impact of the eddy–current shielding on results of BFMR measurements, we study the samples with thicknesses equal to multiples of the technologically meaningful value $d = 10$ nm. Since our investigation is comparative in nature, it is important to characterise samples with well resolved differences in behaviour. Measuring samples with thicker (but still sub–skin–depth) capping layers and comparing them to the reference $d = 10$ nm sample allows us to easily establish the functional dependence of the shielding effect on the layer thickness. For the same reason of maximising measurement accuracy we also use a thick Py layer, although the theory in Ref. [35] predicts a significant impact of eddy currents for much thinner samples, too (see Fig. 6 in Ref. [35]).

The absorption spectrum of the ML is obtained as a ratio of the complex scattering parameter $S_{21}$ for the loaded ML (with a sample on top) to $S_{21}$ of the unloaded ML. Figure 2(a) shows typical absorption spectra obtained using this post–processing procedure for the multilayer with $d = 20$ nm. One sees dips in the absorption spectra taken at different $f$. The amplitude of these dips corresponds to the amplitude of the respective FMR response of the multilayer. Note that at $f = 12$ GHz one also observes a smaller dip at a lower applied magnetic field (~1.1 kOe), which can be identified as the first higher–order standing spin wave mode (1st SSWM).[35]

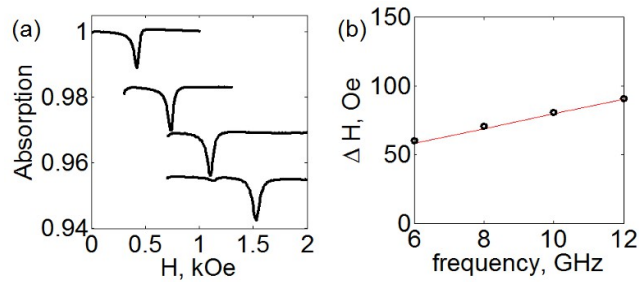

**FIG. 2** (a) Experimental FMR absorption spectra of the multilayer with $d = 20$ nm for $f = 6, 8, 10$ and 12 GHz (from top to bottom respectively). For clarity, each spectrum is vertically offset by – 0.015. (b) Experimental *full width at half maximum* line width $\Delta H$ of the FMR response of the multilayer with $d = 20$ nm as a function of $f$. The straight line is the best fit of the experimental data. Note an extra absorption peak of small amplitude in the lowest trace of Panel (a) (at ~1.1 kOe). This extra peak is the response of the 1st SSWM.

Figure 3 shows the experimental and theoretical amplitudes of the FMR response of the Cu–Py samples. All curves are normalised to the amplitude of the sample with $d$ = 10 nm at 6 GHz. We observe that the amplitude drops very quickly as $d$ is increased. This result confirms our previous theoretical predictions.[36] The decrease in the amplitude is due to eddy currents circulating in the capping layers, which shield the ferromagnetic Py film from the microwave magnetic field [Fig. 1(c)].

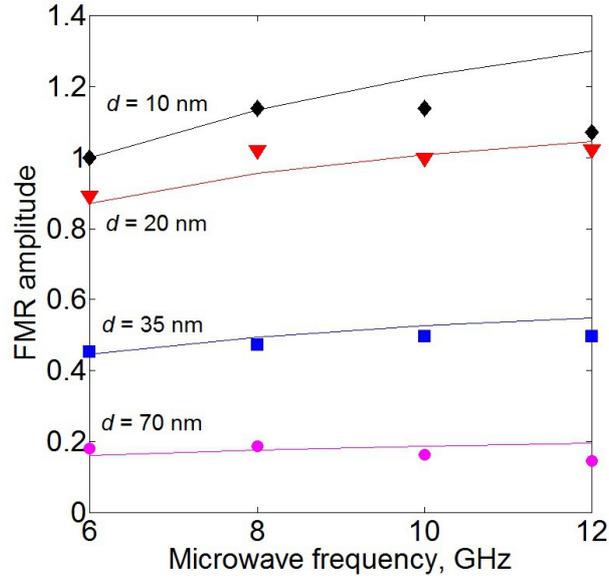

**FIG**. 3 Experimental (symbols) and theoretical (curves) relative amplitudes of the FMR response of the Cu–Py multilayer samples as a function of $f$. All curves are normalised to the absorption amplitude of the sample with $d$ = 10 nm at 6 GHz. Parameters of simulations are from Table 1 and the main text. Some disagreement between the theory and the experiment is attributed to the effects of imprecise sample placement on the microstrip line and the sample dimensions[29] as well as asymmetry of the FMR line shape,[44] which could not be taken into account in calculations.

We also obtain the values of the gyromagnetic ratio $\gamma/(2\pi)$ and of the saturation magnetisation for the FM layer ($4\pi M_s$) by best–fitting experimental data with the Kittel formula[45]

$$f = \frac{|\gamma|}{2\pi}\sqrt{H(H + 4\pi M_s)}. \tag{1}$$

These values are used to extract the values of the Gilbert damping parameter $\alpha_G$. From the measured absorption spectra we extract the *full width at half maximum* line width $\Delta H$ as a function of $f$ and fit the obtained dependence with a straight line[46]

$$\Delta H = \frac{2\alpha f}{|\gamma|/(2\pi)} + \Delta H_0. \qquad (2)$$

The parameters extracted from experimental absorption traces are presented in Table 1.

**Table I**. Parameters extracted from experimental FMR absorption traces for different values of $d$.

| $d$, nm | 10 | 20 | 35 | 70 |
|---|---|---|---|---|
| $\gamma/(2\pi)$, MHz/Oe | 2.998 | 2.996 | 3.025 | 3.030 |
| $4\pi M_s$, G | 8003 | 9033 | 8676 | 8790 |
| $\alpha_G$ | 0.0075 | 0.008 | 0.0086 | 0.0106 |
| $\Delta H(f=10$ GHz$)$, Oe | 90.7 | 80.62 | 100.8 | 120.9 |
| $\Delta H_0$, Oe | 40.84 | 26.44 | 33.71 | 48.95 |

From Table I one sees that we have obtained typical values of $\gamma/(2\pi)$, $4\pi M_s$, and $\alpha_G$ for Py.[2,17–19,22] It is not unusual that the saturation magnetisation for Py samples differs considerably from the standard value $4\pi M_s = 10800$ G. Values in the range from 8000 G to the standard value were found by different authors (see, e.g., Refs. [47–49]). Also, we note that for $d = 10$ nm we find $\alpha_G = 0.0075$, which is slightly lower than but very close to the typical value for Py: 0.008.[19]

The damping parameters for the $d = 10$nm–thick film are somewhat different from the other three samples. Indeed, for the remaining samples one observes a clear trend: an increase in $d$ correlates with an increase in $\alpha_G$ and $\Delta H_0$. Furthermore, this trend possibly suggests that the eddy currents may also contribute to the frequency independent part of the FMR losses $\Delta H_0$. The correlation between $\alpha_G$ and the thickness of the FM film is well–known.[31,38] However, in our experiment the thickness of the FM is constant but the thickness of the NM capping layer is varied.

The clear correlation of the loss parameters with $d$ also suggests that the magnetic quality of the Py layer of the $d = 10$nm–thick film is worse than for the other films. This is seen not only from the large $\Delta H(f=10$ GHz$)$ value for it, but also from the noticeably smaller $4\pi M_s$ than for the other films. This is potentially the reason why the magnetic losses for this film do not follow this trend.

We employ the semi–analytical theory from our Ref. [35] to calculate the FMR amplitude as a function of $d$. We use the data from Table 1 as input parameters. Firstly, we use the bulk Cu conductivity for both NM layers: $\sigma_{bulk} = 5.96 \times 10^7$ S/m. This calculation delivers a $d$–dependence of the FMR amplitude which is significantly steeper than the experimental one.

However, when we assume that the conductivity of the capping layer is $0.4\sigma_{bulk}$ for $d = 70$ nm, 35 nm and 20 nm, and $0.42\sigma_{bulk}$ for $d = 10$ nm, where $\sigma_{bulk} = 5.96 \times 10^7$ S/m is the conductivity of bulk Cu, our result is in reasonably good agreement with the experiment (Fig. 3). Importantly, the simulated raw traces show shapes very similar to ones in Fig. 2(a) – characterised by a very small amplitude of the 1st SSWM with respect to the fundamental mode.

Note that in the second calculation we need to keep the conductivity of the seed Cu layer equal to $\sigma_{bulk}$. Otherwise the calculation delivers much larger relative amplitudes of the 1st SSWM. The increased amplitude of the 1st SSWM is due to strong non–uniformity of the in–plane microwave magnetic field inside the Py layer. The non–uniformity increases with an increase in the eddy–current density inside this layer. The smaller the conductivity of the seed layer, the larger is the eddy current inside the Py layer. Hence, the practically vanishing amplitude of the 1st SSWM in the experimental traces may be considered as an evidence of a noticeably larger conductivity of the seed layer with respect to the capping layer: a large current flowing inside the seed layer makes the microwave eddy–current field in the Py layer more uniform. Indeed, it has been shown experimentally that the electric resistivity of Cu layers strongly depends on the layer on which they grow.[50] More precisely, a Cu layer grown on top of a Py layer may have significantly smaller conductivity than the one grown on a bare substrate.[20]

## 3. Conclusions

We have experimentally investigated the broadband FMR response of metallic magnetic multilayer structures and demonstrated a crucial effect of non–magnetic metallic capping layers of sub-skin–depth thicknesses on the strength of this response. Eddy currents circulating in the capping layers shield the ferromagnetic Py film from the microwave magnetic field. The shielding leads to a strong decrease in the amplitude of the FMR response. This finding has direct implications for microwave characterisation of magnetic materials, including materials for spin–transport applications. They are also important for applications of ferromagnetic films in frequency agile meta–materials. Very often in those experiments, microwave power is incident on just one of the two sample surfaces and is spread over an area at least 100 μm×100 μm in size. These are experimental conditions for which our results are relevant. For instance, these conditions are usually met in the case of the broadband stripline FMR.

**Acknowledgements**

This work was supported by the UPRF scheme of the University of Western Australia and the Australian Research Council.